\documentclass[aps,prd,onecolumn,12pt,preprintnumbers]{revtex4-1}

\usepackage{hyperref}

\usepackage{float}

\usepackage{tikz}

\usepackage{graphicx}
\usepackage{epstopdf}

\usepackage{caption}

\begin{document}

\preprint{HDP: 19 -- 01}

\title{Banjo Ring from Stretching String: A Zero Break Angle Demo}

\author{David Politzer}

\email[]{politzer@theory.caltech.edu}

\homepage[]{http://www.its.caltech.edu/~politzer}

\altaffiliation{\footnotesize Pasadena CA 91125}
\affiliation{}

\date{March 7, 2019}

\begin{figure}[h!]
\includegraphics[width=5.0in]{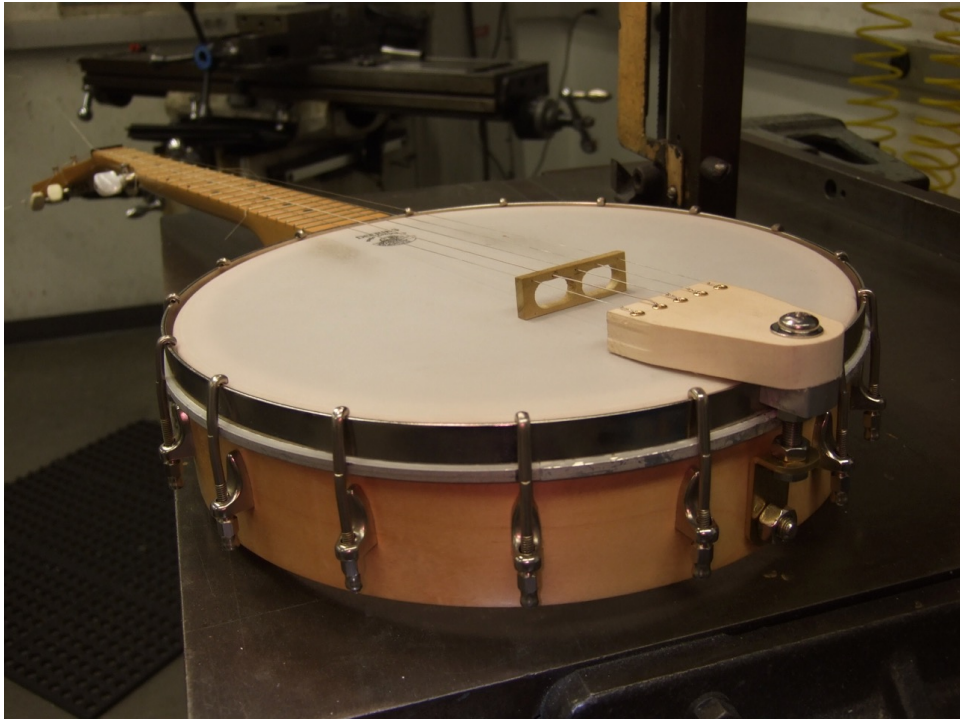}
\caption*{\it  one of several prototypes}\end{figure}

\bigskip

\begin{abstract}
A novel bridge and tailpiece design allows direct comparison of the sound of zero break angle with same banjo (and all its parts) configured to have an angle of $13^{\text{o}}$.  This lends additional support to the 2014 proposal that a key element in banjo sound is the frequency modulation produced by string stretching due to a floating bridge, break angle, and head with substantial motion.  When playing a banjo tune in the $0^{\text{o}}$ configuration, there are enough audio clues that it still sounds like a banjo.  The comparison allows you to judge for yourself to what extent  it's lost its ring or sparkle.

\end{abstract} 

\maketitle{\centerline {\large \bf Banjo Ring from Stretching String: A Zero Break Angle Demo}}

\bigskip

\bigskip

Curiosity about tailpieces and string break angles over the bridge led to an investigation of some of the physics involved.\cite{FM}  A string with fixed ends must stretch when it undergoes transverse vibrations.  However, that stretch is proportional to the square of the vibration amplitude.  Assuming the stretch to be  negligible for small  amplitudes gives rise to a very satisfactory physics description of the vibrating string.  In contrast, the motion of a floating bridge followed by a non-zero break angle produces stretching that is linear in the amplitude.  And that produces tension and frequency modulation which are not necessarily negligible.  The sound of a tone whose modulation frequency is also in the audio range was discovered by John Chowning in the early 1970's.\cite{chowning}  He described it as sounding ``metalilic."

The presence of first order (i.e., linear) string stretching in the banjo is an unequivocal matter of geometry.  The subtle issue is a question of psychoacoutics.  What is it that allows people to identify banjo sound, irrespective of whether the instrument is a bluegrass banjo with lots of metal parts or a 150 year old banjo with not a single scrap of metal?  Banjos were said to ring well before Stephen Foster wrote about them in the 1850's.  The suggestion that banjo ``ring" is an aspect of Chowning's metalilc sound was supported by a variety of examples, demonstrations, and players' observations.\cite{FM}

The details are not repeated here.  Rather, what is presented is the starkest possible comparison: a normal banjo with and without a break angle over the bridge.

\bigskip

\centerline{\bf ZERO DEGREE BREAK ANGLE DESIGN }

\medskip

On a banjo with a normal break angle, the stretching of the strings is first order in the vertical motion of the bridge, i.e., the former is linearly proportional to the latter.   The constant of proportionality goes to zero as the break angle gets smaller.  At zero break angle, the stretching is proportional to the square of the bridge motion.  However, a normal bridge and tailpiece will not work at zero break angle.  The strings would buzz at the bridge as they vibrate upwards; likewise, the bridge would buzz on the head.

The design used here obviates those problems.  It also allows a sound comparison with the same banjo, bridge, and tailpiece but with a break angle of $13^{\text{o}}$  --- simply by flipping the tailpiece over and reinstalling the strings.

\begin{figure}[h!]
\includegraphics[width=6.0in]{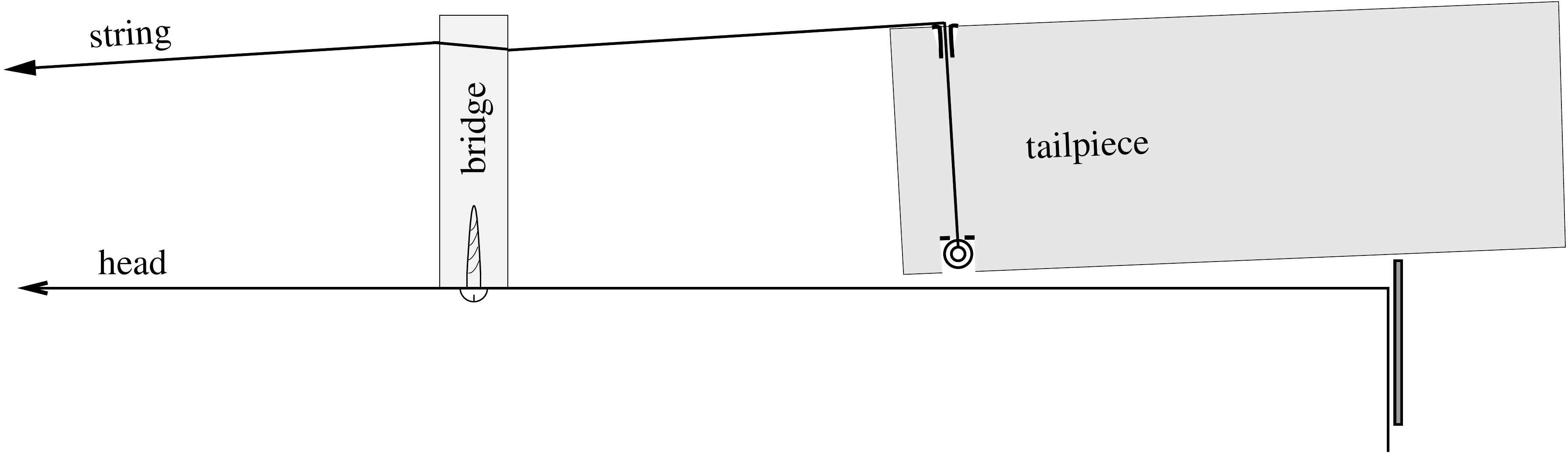}
\caption{Bridge-tailpiece schematic -- not to scale}\end{figure}

The strings pass through holes in the bridge.  The hole diameters are all about 10\% larger than the strings.  The holes angle down towards the tailpiece at $5^{\text{o}}$ relative to the head.  The strings enter the top of the tailpiece through ferrules.  They are secured by their ball ends in steel-washer-lined holes in the bottom.  The $5^{\text{o}}$ means that the centerlines of the holes drop $0.0135''$ as they traverse the $0.154''$ thickness of the bridge.  The string gauges are 10-12-14-22w-10.  So the oversize holes and the $5^{\text{o}}$ drop ensure that strings sit snugly on the bottoms of the holes (much like they normally sit in slots on the top of a bridge) and do not bind in some awkward way in the hole.  Even when the string vibration motion is upward, the $5^{\text{o}}$ break is sufficient to keep it in contact with the bottom of its hole.  (That's clear because $5^{\text{o}}$ is perfectly adequate on a normal Old-Time banjo to keep it from buzzing.)  The hole height was chosen to provide the same action as a normal $5/8''$ bridge and a Presto-style tailpiece.  (Note that, with zero break angle, the head is perfectly flat, even with the strings at full tension.)

The bridge is glued to the smooth side of a bottom-frosted head and further secured with three brass \#1 -- $1/4''$ screws.  (I didn't find any glue that adhered to the mylar, even after sanding; so I used contact cement.)  The final weight of the bridge, including screws, is 2.6 gm (i.e., pretty standard for a steel string and mylar head set-up).

The tailpiece is designed so that only very minor adjustment is needed, when under full string tension, to bring the bridge-to-tail strings into parallel with the bridge-to-nut strings.  The slight off-set due to the $5^{\text{o}}$ channels has no impact on string stretching.  In particular, the stretch due to bridge motion is precisely the same as would occur for upward bridge motion with strings going straight over a normal bridge with no break angle.  The wood of the tailpiece is 35 gm, which is relatively hefty --- chosen to minimize its motion.  Including aluminum spacer and steel screw, washer, and three nuts, the tailpiece is 63 gm.

\newpage

\begin{figure}[h!]
\includegraphics[width=5.0in]{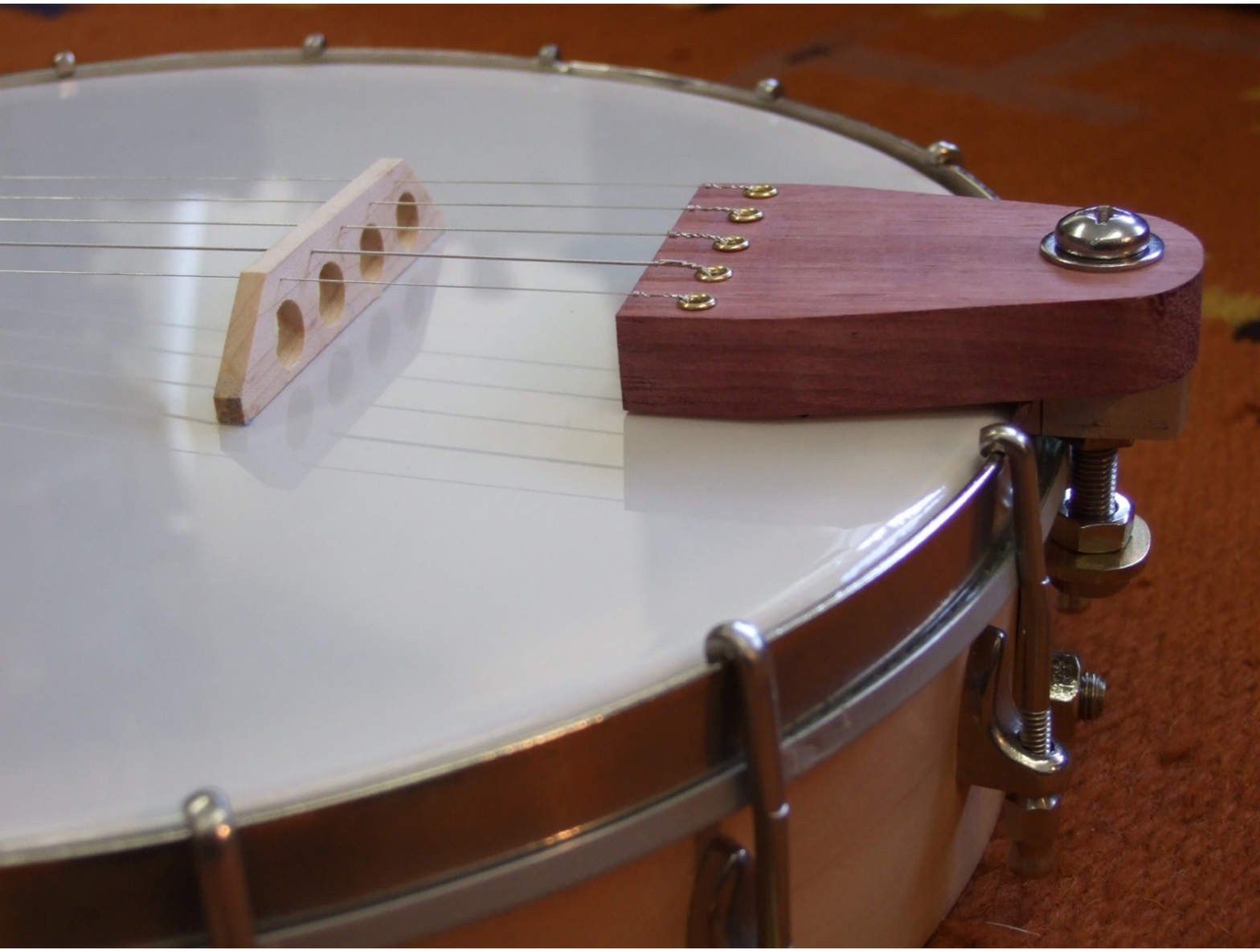}
\includegraphics[width=5.0in]{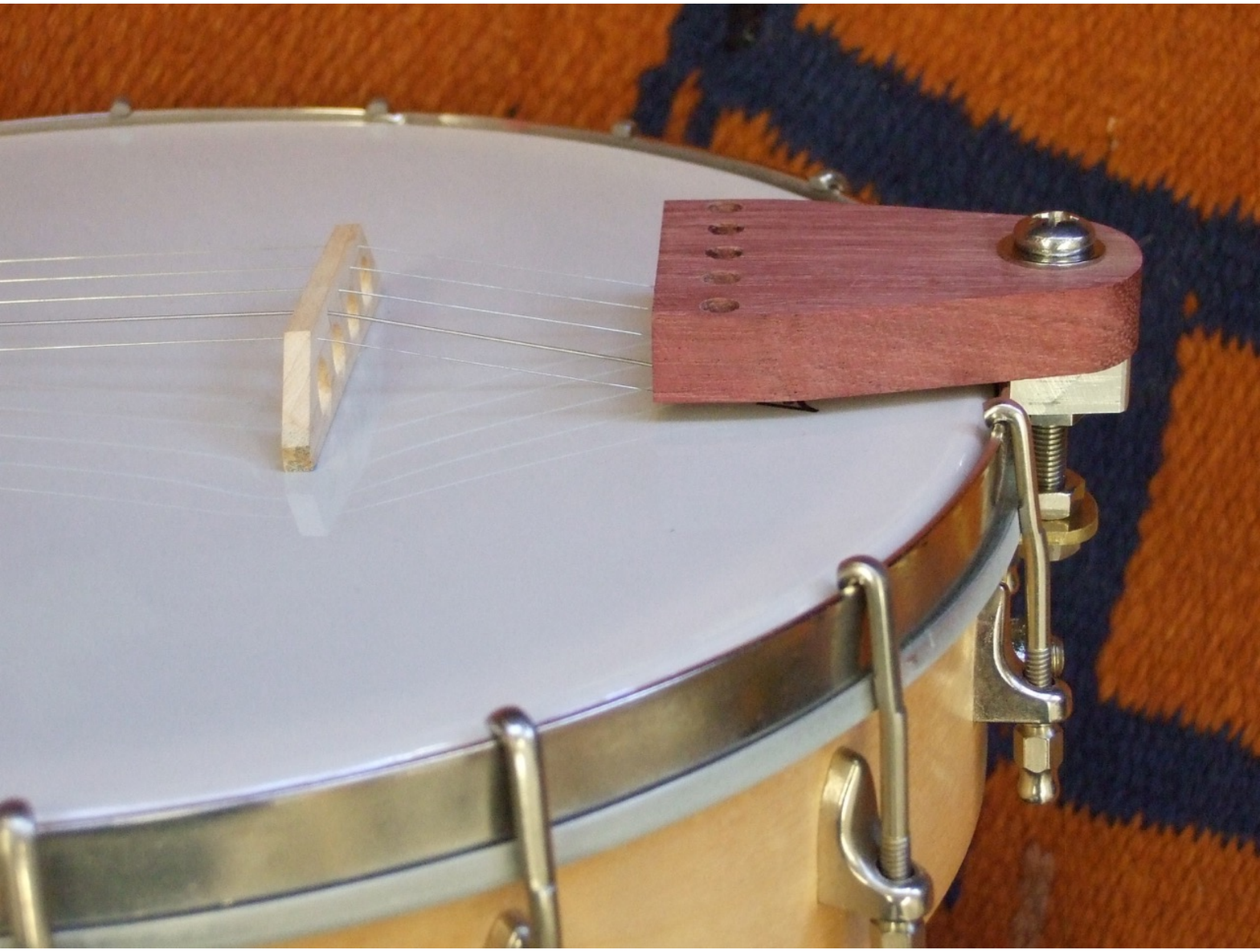}
\caption{The bridge and tailpiece described above and used in the sound samples with zero and $13^{\text{o}}$ break angle; the woods are maple and purpleheart, respectively.}
\end{figure}

\newpage

\centerline{\bf A COUPLE OF TUNES FOR COMPARISON}

\medskip

The banjo is a year 2000 Deering Goodtime, played with a solid disk back spaced $1/4''$ off the rim.  The head was tensioned to 91 on a DrumDial.  For two tunes, each played in the two configurations:

\href{http://www.its.caltech.edu/~politzer/zero-break/A.mp3}{click here for sample A}, 
 
\href{http://www.its.caltech.edu/~politzer/zero-break/B.mp3}{click here for sample B},
 
\href{http://www.its.caltech.edu/~politzer/zero-break/C.mp3}{click here for sample C}, and
 
\href{http://www.its.caltech.edu/~politzer/zero-break/D.mp3}{click here for sample D,}
 
\noindent or go to http://www.its.caltech.edu/\url{~}politzer/zero-break/A.mp3 and then switch A to B, C, and D.  If you can't hear which is which, this exercise is pretty bootless.

\bigskip

\centerline{\bf DISCUSSION}

\medskip

Audacity$^{\tiny{\textregistered}}$'s frequency analyses of the finger-picked tune are displayed in FIG.s~3, 4, \& 5.  These serve as a guide to what you're hearing.  All graphs show that there is little difference between the tailpiece set-ups below 1500 Hz.  That range includes all the fundamental frequencies of the notes played and many of their first few harmonics.  (The fundamental frequencies of those notes range from 131 Hz [open $4^{\text{th}}$] to 1047 Hz [$12^{\text{th}}$ fret $1^{\text{st}}$ string].)  And it accounts for virtually all of the sound energy.  So the two set-ups are equally loud.  But the $0^{\text{o}}$ version is generally much weaker above 1500 Hz.  (The next section, {\bf \small SINGLE STRING PLUCKS}, takes a closer look at the time evolution of individual notes.)

\begin{figure}[h!]
\includegraphics[width=6.5in]{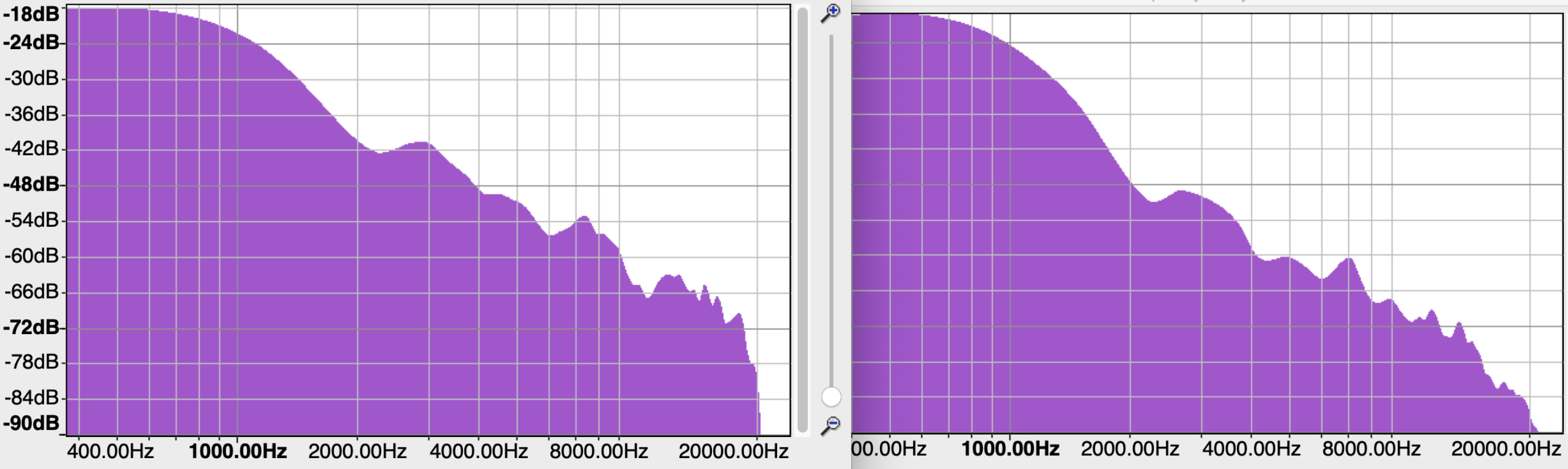}
\caption{low resolution spectra for the full 34 seconds of the picked tune --- $13^{\text{o}}$ on the left, \& zero on the right}
\end{figure}
\begin{figure}[h!]
\includegraphics[width=6.5in]{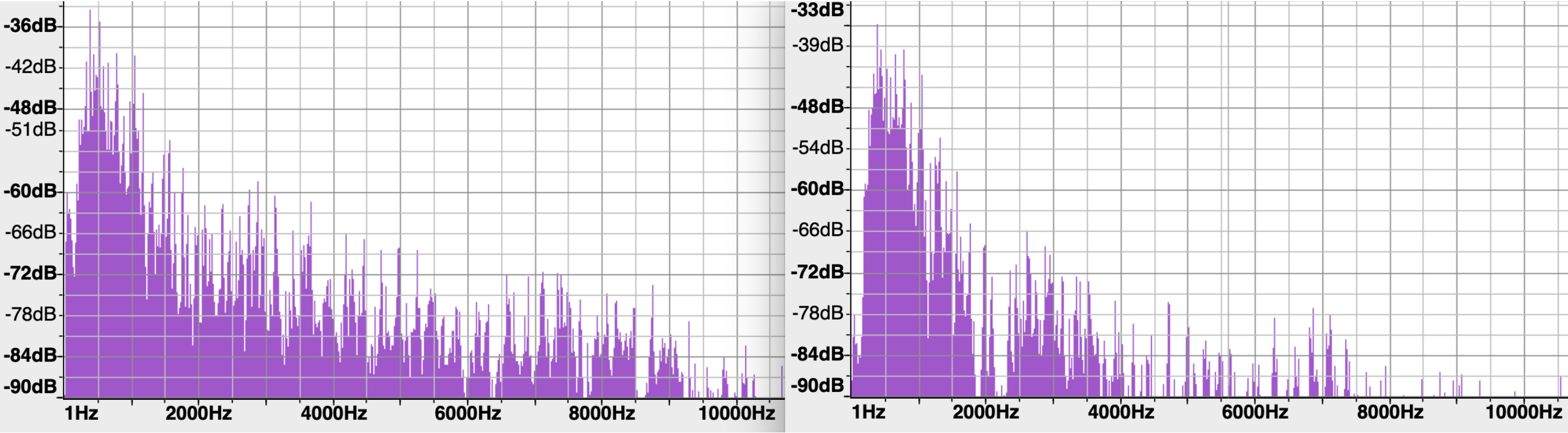}
\caption{high resolution version of FIG.~3}
\end{figure}

\begin{figure}[h!]
\includegraphics[width=6.5in]{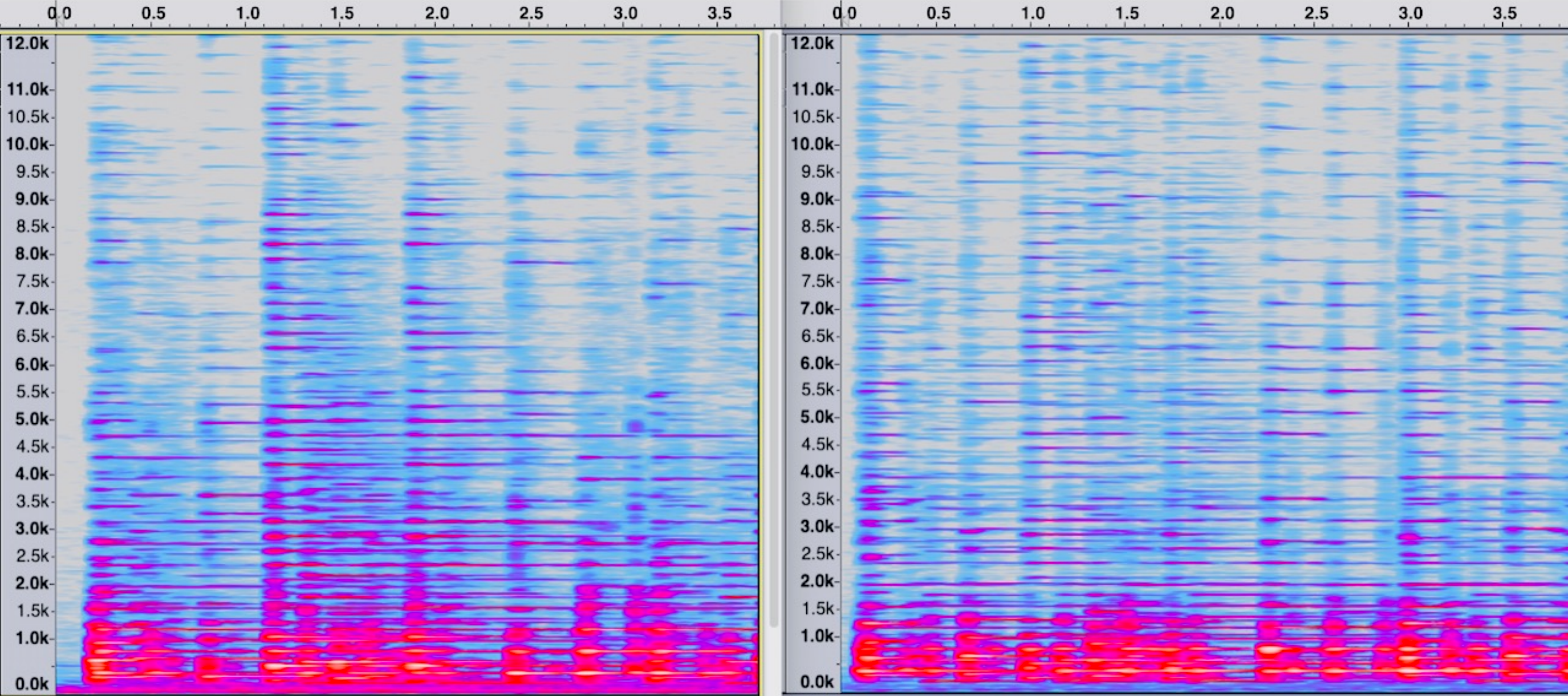}
\caption{spectrograms of the first few seconds of the picked selection, $13^{\text{o}}$ (left) and $0^{\text{o}}$ (right), Hz. {\it vs.} seconds}
\end{figure}

(Note on spectrograms: the choice of frequency resolution, $\Delta \nu$, produces a time resolution, $\Delta t$, subject to the relation  $\Delta \nu \Delta t \approx 1$, i.e., the ``uncertainty relation." Hence, high resolution in frequency necessarily smears out the results in time.)

Nevertheless, both versions sound like a banjo.  (What did you expect?  A French horn or an oboe?)

The mental identification of ``banjoness" (performed effortlessly and unconsciously, at least by someone who has listened to a significant amount of banjo music) is, presumably, based on many clues.  Facial and voice recognition of a particular person certainly work that way.  It's likely that no one clue is absolutely essential.  Here are a couple of examples from the far corners of banjo playing that illustrate how perceptions can be confused.  The Reverend Gary Davis occasionally picked up a 6-string banjo and even recorded some on one.  He played the same repertoire and in exactly the same style as his guitar playing.  It's easy to miss any difference.  Adrian Legg's compositions include a few tunes and arrangements that are immediately recognizable as banjo music.  But it's all played on guitar (admittedly with a healthy dose of real-time electronic effects).  Harvey Reid sometimes plays a 6-string banjo.  Even straight acoustically, he can make it sound like almost any plucked string instrument, e.g., from his ``faux frailing" banjo sound to a whining, driving solid body guitar.

The other obvious physical feature common to all banjos is the head.  A mylar or skin head sounds nothing like a wooden head.  Instruments with the latter sound like dulcimers, irrespective of their appearance or how they're played.  In ref.~\cite{FM}, I argued that, under the action of similar strings, a proper drum head would move through much greater distance than a wood sound board.  That obviously implies that the string energy is turned into sound faster with the drum head.  So a normal banjo is louder but for a shorter period.  But the timbre changes, too.  Part of that could be more high frequency dissipation with the wood head while the vibrational energy is still in the string, i.e., before being converted into sound.  But there's also the fact that the string is stretching much more while it is vibrating with the mylar or skin head.  On the other hand, a proper acoustician would point out that a drum head is reasonably approximated as a ``membrane," whose restoring force is principally due to tension, while a sound board is properly described as a thin plate, whose restoring force is stiffness.  The spectra of resonant frequencies of the two are rather different.

Ref.~\cite{FM}, contains explicit examples of the sound of frequency modulation, both generated analog, i.e., by manually moving the bridge up and down and then speeding up the whole recording, and digitally, i.e., computer generated files of sinusoidally modulated sinusoidal functions.  In all cases, there is a discernible ring.  But the actual sound of a played banjo is quite complex.  Many things are going on at once.  The bridge rocks and goes up and down under the influence of all the strings.  And that, in turn, effects the motion of all the strings.  So the strings certainly influence the motion of each other.

Chowning's insight was that frequency modulation could be the basis of a very rich range of synthesis possibilities.  In fact, that's how and why it became the basis of Stanford University's second biggest money-making patent.  It was licensed to Yamaha. Jointly with Stanford, they developed the DX7.  Released in 1983, it was the first commercially successful electronic keyboard synthesizer.  The point is that FM synthesis can easily get you almost anywhere in timbre space.  So it is not clear that successfully synthesizing banjo sound using frequency modulation would add any additional support to the notion that break angle and stretching are the essential elements.  Chowning himself began with the hitherto vexing brass instrument sound.  With an appropriately chosen amplitude envelope function and frequency modulation of the note, he found he could make plausible imitations of brasses, from tubas to cornets.  (Note added:  A colleague pointed out that frequency modulation can be added to an otherwise generic plucked-string-sound synthesizer to get plausible banjo-like sounds.\cite{K-S})

The logic of the identification of frequency modulation as the source of ``ring" is, therefore, necessarily a bit indirect.  As is well-accepted among players, steeper break angle makes the sound ``ringier."  In the comparisons presented here, the only physical change between banjos is the dramatic change in break angle.  From a physics perspective, one must ask, ``What is the relevant change in the sound producing system as a result of that angle?"  Steeper angles do not improve the motion transfer from the strings to the bridge (discussed in {\large $\S$}{\bf \small TAILPIECE DOWN PRESSURE} below).  The only consequence I can identify in the equations of motion for the system is the increased stretching of the strings as the floating bridge moves up and down.  This effect is a continuous function of angle.  In words (rather than in math concepts), the break angle determines the amount of stretching, and the stretching is smallest at zero angle.

\bigskip

\centerline{\bf SINGLE STRING PLUCKS}

\medskip

Listening to and analyzing the pluck of a single string offers further insight into the time evolution of the sound.  Among the many possibilities, I found the plucks of the open $4^{\text{th}}$ string the most revealing.  Here, the other strings are left open (undamped).  I chose one typical pluck for each tailpiece configuration.  In the following example, the first (and left) is the $13^{\text{o}}$ break angle; the second (and right) is $0^{\text{o}}$.

Listen to the comparison cut off at 2.5 seconds after the pluck:
\href{http://www.its.caltech.edu/~politzer/zero-break/4th-comparison-2.5sec.mp3}{Click here for a comparison of $4^{\text{th}}$ string plucks} or get \newline
http://www.its.caltech.edu/\url{~}politzer/zero-break/4th-comparison-2.5sec.mp3.

Here are the same two plucks, this time cut off after 0.5 seconds:
\href{http://www.its.caltech.edu/~politzer/zero-break/4th-comparison-0.5sec.mp3}{Click here for a  comparison of $4^{\text{th}}$ string plucks} or get \newline 
http://www.its.caltech.edu/\url{~}politzer/zero-break/4th-comparison-0.5sec.mp3.

Spectrograms of these two versions are shown in FIG.s~6 and 7.  (The pile-up in FIG.~7 at the end of the 0.5  second selections is an artifact of the math algorithm and the sharp cut-off of the sound file.)

\begin{figure}[h!]
\includegraphics[width=5.0in]{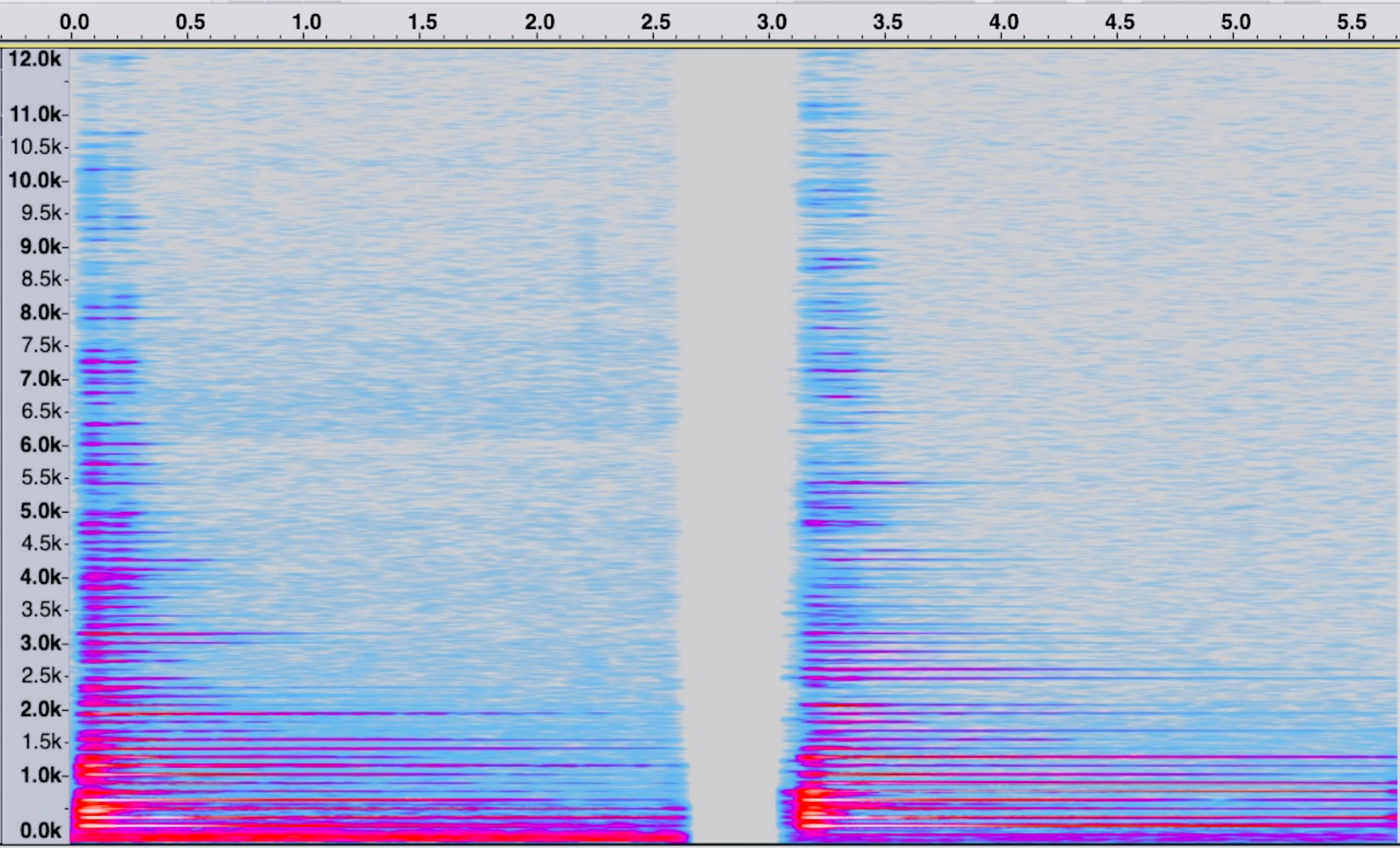}
\caption{High frequency resolution spectrograms of single plucks of the $4^{\text{th}}$ string --- $13^{\text{o}}$ and then $0^{\text{o}}$, both cut off after 2.5 seconds. }
\end{figure}

\begin{figure}[h!]
\includegraphics[width=5.0in]{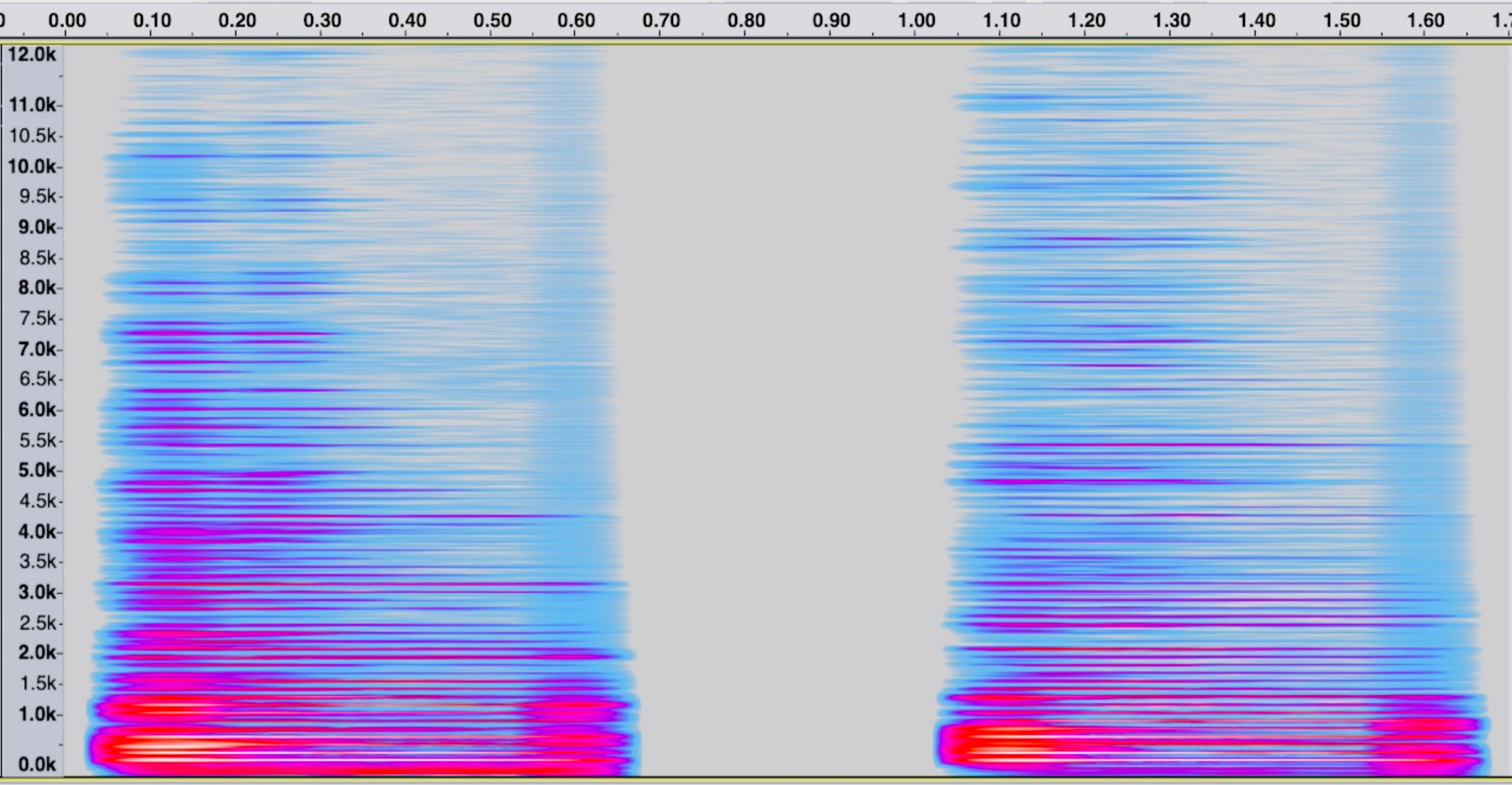}
\caption{High frequency resolution spectrograms of single plucks of the $4^{\text{th}}$ string --- $13^{\text{o}}$ and then $0^{\text{o}}$, both cut off after 0.5 seconds. }
\end{figure}

Again, there is little difference below 1500 Hz.

The effect seems strongest with the 4th (lowest) string.  This agrees with the ``theory."  The stretching is proportional to bridge motion, and that is greatest for the lowest the notes.  
The dramatic differences between the two set-ups are confined to the first 0.2 - 0.3 seconds of the pluck.  The longer sustain is dominated by the common sound of lower harmonics.  On the other hand, in normal playing, notes typically come at 4 per second or faster, often more than one at a time.  So there are lots of things are going on at once.

\bigskip

\centerline{\bf TAILPIECE DOWN PRESSURE \& STRING DRIVING EFFICIENCY}

\medskip 

Some people believe that the down pressure due to break angle contributes to effective transmission of string motion to the head, and the more the better --- at least up to a point.  However, the zero break configuration was perfectly loud.  Furthermore, a simple physics analysis suggests just the opposite: an increase in equilibrium down pressure reduces head motion in response to a given string motion.  And that's what ultimately chokes out the sound at extreme break angles.

At equilibrium, when the strings are at rest, their downward force on the bridge is exactly canceled by the upward force from the deformed head.  That is a {\it stable} equilibrium.  If the bridge is displaced upward,  the combined forces of the strings (still not vibrating themselves) and the head, push it backward.  The string downward force increases for two reasons.  The tension increases because the strings are stretched, and the break angle becomes steeper (so the downward component of tension is greater).  Conversely, the head upward force decreases when the bridge is displaced up (for exactly the analogous reasons).  So an upward bridge displacement results in a greater {\it net} downward force on the bridge from strings and head.

The important point is that the restoring force (i.e., the force tending to return the bridge to its equilibrium position) for a given displacement is greater for a larger equilibrium break angle.  The string stretching and increase in tension are greater, and the downward component of the tension (even for a given value of tension) is greater.

In the context of normal playing, the head moves because string vibrations make small changes in the down force of the strings on the bridge.  However, the bridge moves under this force {\it and} the net force from its being displaced.  Hence, increased break angle {\it reduces} the net motion of the bridge.  Ultimately, it can be enough to produced a discernibly choked response.

At the other end, a zero break angle minimizes the return force of quiet strings and head on the bridge.  For small amplitudes, bridge motion is solely determined by the vibrating string forces.  Consequently, the head moves more air for a given amplitude pluck.  However, the perception of loudness is not purely an issue of sonic power.  There is a frequency dependence, too.  I believe that we are more likely to take note of higher frequencies.  So tiny increases in the energy of high frequencies will produce a perception of louder sound.

\bigskip


\centerline{\bf CONTRIBUTORS TO BANJO SOUND}

\medskip

Virtually every design aspect contributes in some way to the sound of a banjo.  And some design choices emphasize what some people consider ``banjoness."  Head density and tension are obvious examples.  I've written previously about a few not-so-obvious examples. Bridge mass has a huge impact.\cite{bridges}  Extreme added-on mass is one common form of mute.  It comes with dramatic changes in timbre.  In particular, the sustain is increased, and higher overtones are suppressed relative to lower ones.  But these effects are clearly discernible even within what is considered the normal range for bridge masses.  The geometry at the edge inside the pot and just below the head also can be chosen to impact the the amount of ``ping" associated with each struck note.\cite{ping}  And, of course, resonator backs emphasize the attack and very high overtones.\cite{backs}

But there's something that makes all (real) banjos recognizable as banjos.  And ``all" includes low-tuned, $13''$, fretless, skin-head open-backs constructed as they were in the mid-$19^{\text{th}}$ Century.  The combination of first order string stretching and consequent frequency modulation is one of the few (though not only) things they have in common.  So maybe it really is the key.

\end{document}